\documentclass[12pt,oneside]{article}
\input psfig
\begin{document}
\title{\bf A spacetime dual to the NUT spacetime}
\author{Mohammad Nouri-Zonoz$^1$ \thanks{Email: mnzonoz@ast.cam.ac.uk}
 , Naresh Dadhich$^2$ \& D. Lynden-Bell$^1$\\
 {\it $^1$Institute of Astronomy, Madingley Road, Cambridge. CB3 0HA}\\
{\it $^2$IUCAA, Post Bag 4, Pune University Campus, Pune, 411 007, India.}}
\maketitle
\begin{abstract}
By  decomposing the Riemann curvature into electric and  magnetic
parts,  a duality transformation, which involves  interchange  of
active and passive electric parts, has recently been proposed. It
was shown that the Schwarzschild solution is dual to the one
that describes the Schwarzschild particle with cloud of string dust 
or a global monopole. Following  the same procedure 
we obtain the solution  dual
to the NUT spacetime.
\end {abstract}

\section{Introduction}
In  analogy  with  the  electromagnetic  field  as  well  as  the
resolution of the Weyl curvature, we decompose the entire Riemann
curvature  components  relative to a timelike  unit  vector  into
electric and magnetic parts (Dadhich 1997). The projection of
Riemann tensor and its
double  dual are respectively identified as the active and passive electric
parts, while  the projection of the single dual is the magnetic part. The
electromagnetic parts are the second rank 3-tensors orthogonal  to
the  timelike resolving vector, the electric parts are  symmetric
while  magnetic part is trace-free and consists of the  symmetric
Weyl magnetic part and an antisymmetric part representing  energy
flux.
We  can now write the vacuum Einstein equations entirely in  terms
of  electromagnetic parts. It is in general symmetric in
active  and passive electric parts. The duality transformation we wish to
consider  is  the  interchange of active  and  passive  parts, which
would in general imply interchange of the Ricci and the Einstein tensors.
This is becuase the former results from contraction of the Riemann while
the latter from its double dual. The  vacuum equation is obviously
duality invariant. However it turns out
that for
the interesting vacuum solutions it is possible to break the  symmetry
and  yet obtain the characteristic vacuum solution. This
 is so for all the black hole solutions (Dadhich 1997) and as we  now 
show for the  NUT solution. What really happens
is that in obtaining the vacuum solution there always remains one extra
equation corresponding to the Laplace equation . That it is
implied by the others. Now throwing in some appropriate distribution
on the right of it will not disturb the vacuum solution but it will
now make the equation duality non-invariant. The solution of the dual
equation gives a metric which can be interpreted as the vacuum spacetime 
with a global monopole. Barriola and Vilenkin (1989) have already obtained
what can be called the spacetime of the Schwarzschild black hole with 
a global monopole (well outside the core of the monopole) . It is shown that
this is dual to the Schwarzschild solution (Dadhich 1997). Similarly
the solution dual to the Kerr solution has also been constructed (Dadhich \&
Patel 1998a). \\
In the case of non-empty space, the duality transformation would
imply $T^i_k \rightarrow T^i_k - (1/2)Tg^i_k$, and in particular a
fluid solution maps into a fluid solution with $\rho \rightarrow
(\rho + 3p)/2$ and $p \rightarrow (\rho - p)/2$, and heat flux and
pressure anisotropy remaining unaltered (Dadhich, Patel \& Tikekar
1998c). It then follows that the stiff fluid is dual to dust, the
radiation and the de Sitter models are self-dual, and perfect fluid
with the equation of state $\rho + 3p = 0$ is dual to flat
spacetime. \\
Note this duality transformation is different from Ehlers' duality
transformation (Ehlers 1962, Geroch 1971). The former  interchanges
the active and passive electric parts of the Riemann tensor  and
generates a global monopole. On the other hand the latter
amounts to generating a symmetric magnetic part in the otherwise gravomagnetic
part-free Schwarzschild solution to give the NUT space, which
has a gravomagnetic monopole (Geroch 1971, Lynden-Bell \& Nouri-Zonoz 1998).\\
In the next section we formulate the alternate vacuum equation in
terms of the electromagnetic parts of the field and obtain the solution dual
to  NUT. This will be followed in section 3 by studying
geodesics of the dual solution to bring out the effect of its global monopole
charge, and we finally conclude with a discussion. \\
\section {Dual-NUT solution}
We resolve  the Riemann curvature relative to  a  timelike  unit
vector as follows:
$$E_{ac}=R_{abcd}u^b u^d \;\;\;\;\;\;\;\; , \;\;\;\;\;\;\;\;
{\tilde E}_{ac} =*R*_{abcd}u^b u^d \eqno(1)$$

$$H_{ac}= *R_{abcd}u^b u^d = H_{(ac)} + H_{[ac]} \eqno(2)$$
 where

 $$H_{(ac)} = *C_{abcd}u^b u^d \eqno(3)$$
 $$H_{[ac]} ={1\over 2} \eta _{abce} R^e_d u^b u^d. \eqno(4)$$
 Here $C_{abcd}$ is the Weyl conformal curvature, $\eta _{abcd}$ is the
 4-dimensional volume element. The following relations are also satisfied,
 $$ E_{ab}=E_{ba} \;\; ,\;\; {\tilde E}_{ab}={\tilde {E}}_{ba}\;\; ,\;\;
  ( E_{ab}, {\tilde {E}}_{ab},H_{ab}) u^b = 0 $$ and
 $$ H=H^a_a=0 , u^a u_a=1.$$
In (Dadhich 1997), $E_{ab}$ and ${\tilde E}_{ab}$ are respectively termed as 
active and passive parts. The former refers to $R_{0\alpha 0 \alpha}$ and 
the latter to $R_{\alpha \beta \alpha \beta}$ ,  $\alpha , \beta = 1,2,3$
components of curvature. It can be argued (Dadhich 1997b) that distribution 
of matter-energy acts as gravitational charge for $E_{ab}$ while gravitational 
field energy produces space curvature and acts as charge for 
${\tilde E}_{ab}$. In the context of the Schwarzschild field, it can be 
shown that the active part is derived from matter-energy and the 
passive part from the gravitational field energy.  
Using the above decomposition one can write the Ricci tensor in 
the following form 
$$R_a^b=E_a^b +{\tilde {E}}^b_a+ (E + {\tilde {E}})u_a u^b -
{\tilde {E}}g^b_a+{1\over 2}
(\eta _{amnc} H^{mn}u^b u^c + \eta ^{bmnc}H_{mn}u_a u_c) \eqno (5)$$
 where $E=E^a_a$ and ${\tilde {E}}={\tilde {E}}^a_a$.
 We define $E={\tilde {E}}
 -{T\over 2}$  to be the gravitational charge density while
 $\tilde {E}=T_{ab}u^a u^b$ defines the
 energy density relative to the unit timelike vector
 $u^a$. The vacuum equation, $R_{ab}=0$ is in general equivalent to
 $$ E \; {\rm or} \; {\tilde {E}}=0 \;\;\; ,\;\;\; H_{[ab]}=0
 =E_{ab}+{\tilde {E}}_{ab}. \eqno(6)$$
 This set is symmetric in $E_{ab}$ and ${\tilde {E}}_{ab}$.\\
Consider the set
 $$H_{[ab]}=0={\tilde {E}} \;\;\;\;\; , \;\;\;\;\; E_{ab}+{\tilde {E}}_{ab}
 =\lambda w_a w_b \eqno(7)$$
 where  $w_a$ is the unit spacelike vector orthogonal to $u_a$ and
$\lambda$ is a scalar function to be determined from the solution.
 Clearly this set  is not
 symmetric in $E$ and $\tilde E$, but it will still yield
Schwarzschild  as the  general  solution for spherical symmetry 
(Dadhich, 1997).
Not only that, it  yields the Kerr solution as well, and we now
 show that it gives the NUT solution. Writing the metric in the
following  form (Newman et al 1963),
 $${ds^2}=A{(dt-2l cos \theta d\phi)}^{2}-B{dr^2}-(r^2+l^2)
 (d\theta^2+sin^2\theta d\phi^2),\eqno (8)$$
 where $A$ and $B$ are functions of $r$ and $t$ only. We wish to find
 the solution of equation (7) for the above metric.
 Now the equation $H_{[ab]} = 0$ will lead to the time independence
 of the solution,
while $(E +{\tilde {E}} )^2_2 = (E +{\tilde {E}} )^3_3 = 0$  gives
          $$ AB = 1 . \eqno(9)$$
Substituting this in ${\tilde {E}} = 0$ leads to the NUT solution,
$$A = B^{-1} =1-{2(mr+l^2)\over(r^2+l^2)} .\eqno(10) $$
Note that we did not use the extra equation $(E + \tilde E)^1_1 = -
\lambda$ which in this case gives  $\lambda = 0$. This is why the
introduction of $\lambda$ term on the right of eqn.(7) did not affect the
vacuum solution and this happens for all stationary solutions including
electrovac ones. \\
Now we define the duality transformation by

 $$E_{ab}\leftrightarrow {\tilde {E}}_{ab} \;\;\;\;\; ,
 \;\;\;\;\; H_{ab} \leftrightarrow - H_{ab} \eqno(11)$$
 which implies $R_{ab} \leftrightarrow G_{ab}$, because contraction of
 Riemann
 tensor is Ricci tensor while that of its double dual is Einstein tensor
 (Misner et al. 1973).
 Equation (7) then transforms into
 $$H_{[ab]}=0=E \;\;\;\;\; , \;\;\;\;\; E_{ab}+{\tilde {E}}_{ab}
 =\lambda w_a w_b . \eqno(12)$$
 The general solution of this set of equations turns out to be

 $$A = B^{-1} = C\left[ D-2l^2/(r^2+l^2)\right] + {F \over \sqrt{r^2+l^2}}
 \left[ D-l^2/(r^2+l^2)\right] ^{1/2}\eqno(13a) $$
 where $C$, $D$ and $F$ are constants. One can put $D=1$ without loss
 of generality because it corresponds to the rescaling of the time
 coordinate
 and the  NUT parameter $l$. To find the constants $F$ and $C$ we use  
the general solution of equation (12) for a spherically symmetric spacetime 
which has the following form (Dadhich 1997)
 $$g_{00}=-g_{rr}^{-1}=1-2k-2m/r \eqno(13b)$$
where $k$ is a constant.
There are two different interpretations for the above metric. In the first
one it has been interpreted as the spacetime associated with a particle of
mass $m$ centered at the origin of the system of coordinates surrounded by a 
spherical cloud of strings of gauge invariant density $2k\over r^2$ 
(Letelier 1979). In another interpretation it is shown to be an approximate
solution of the Einstein equations for the spacetime outside
a global monoploe which has been formed as a
result of a global $O(3)$ symmetry breaking into $U(1)$ 
(Barriola \& Vilenkin, 1989).
In the rest of this paper we will stick to the  interpretation in
terms of the global monopole and elaborate more on it. Now going back to the
determination of the constants $F$ and $C$ we note that one expects (13a)
to reduce to (13b) when  $l=0$ and as a result  $C=1-2k$ and $F=-2m$. 
So we have
 $$A = B^{-1} = \left[ 1-2k-2{mr+l^2(1-2k)\over (r^2+l^2)}\right]
. \eqno(13c)$$
 Again we have $AB = 1$, because the  equation
 that  yielded this remains unaltered, while $E = 0$ then gives  the
 equation
 $$A^{\prime\prime} + 2rA^\prime/(r^2 + l^2) + 4l^2 A/(r^2 + l^2)^2
 = 0\eqno(14)$$
 whose general solution is (13a). Equation (13c) is the spacetime 
dual to the NUT spacetime. Putting $k = 0$ one obtains NUT itself.
The energy momentum distribution implied by  solution (13c)  is

 $$ G^0_0 = G^1_1 = {2k\over (r^2 + l^2)} \eqno(15)$$
 which reduces to that of the Schwarzschild global monopole when $ l = 0$.
Global  monopoles  are supposed to be created as a result
of global symmetry breaking in phase  transitions
 in the early Universe ( Vilenkin \& Shellard 1994). The simplest
radially symmetric global monopole is modelled  by  a  triplet scalar,
 $$\phi ^a = \eta f(r)x^a/r \eqno(16)$$  where
 $x^a x^a = r^2$ with the Lagrangian,
 $$L={1\over 2}\partial _\mu \phi^a \partial ^\mu \phi^a - {1\over
 4}\lambda
 (\phi^a \phi^a - \eta ^2)^2 \eqno(17)$$
This model has a global $O(3)$ symmetry which is
spontaneously broken  to $U(1)$.
At large $r$ outside the monopole core, 
 where $f = 1$, it would generate the stresses  as  given
 by (15) with $l = 0$ i.e., outside the monopole core, the 
energy-momentum tensor can be approximated as
$$T^0_0\approx T^1_1 \approx \eta^2/r^2 \;\;\;\;\;\;\;\;\;\; 
T^2_2=T^3_3 \approx 0 $$
The general solution of the Einstein equations with this energy-momentum
tensor is given by (13b) where $2k=8\pi G \eta^2$ (Barriola \& Vilenkin, 1989).
 
We now show that in the NUT case the scalar field configuration
corresponding to (16) is
 $$\phi ^a = \eta f(r){x^a\over  \sqrt{l^2+r^2}} \eqno(18)$$
 where $x^a x^a = r^2 +l^2$ and
 $$x^1=\sqrt{l^2+r^2}\;\; {\rm sin}\theta{\rm cos}\phi$$
 $$x^2=\sqrt{l^2+r^2}\;\; {\rm sin}\theta{\rm sin}\phi \eqno(19)$$
 $$x^3=\sqrt{l^2+r^2}\;\; {\rm cos}\theta.$$
 Now using metric (8) we can write the Lagrangian (17) for  the scalar
 field $\phi$ (18) in the following form
 $$L={\eta ^2 {f^\prime}^2\over 2B}+{\eta ^2 f^2\over (r^2+l^2)}-
 {\lambda \over 4} \eta ^4 (f^2-1)^2\eqno(20)$$
 where $f^\prime = \partial_r f$. The equation of motion for the field
$\phi$ will be given by
$$B^{-1}[f^{\prime\prime} + f^\prime (\frac{A^{\prime}}{2A} -
\frac{B^{\prime}}{2B} + \frac{2}{r})] - \frac{2f}{r^2} - \lambda \eta^2
f(f^2 - 1) = 0 .\eqno(21)$$
This admits an approximate solution $f = 1$ for large $r$ when O($r^{-2})$
is ignorable.\\

Then $T_{\mu\nu}=2{\partial L
 \over \partial g_{\mu\nu}}-L g_{\mu\nu}$ leads to
 $$T^0_0={\eta ^2 {f^\prime}^2\over 2B}+{\eta ^2 f^2\over (r^2+l^2)}+
 {\lambda \over 4} \eta ^4 (f^2-1)^2$$

 $$T^1_1=-{\eta ^2 {f^\prime}^2\over 2B}+{\eta ^2 f^2\over (r^2+l^2)}+
 {\lambda \over 4} \eta ^4 (f^2-1)^2 \eqno(22)$$

 $$T^2_2=T^3_3={\eta ^2 {f^\prime}^2\over 2B}+
 {\lambda \over 4} \eta ^4 (f^2-1)^2 .$$
 Now for $r\rightarrow \infty$ and $f=1$ we get
 $$T^0_0=T^1_1={\eta^2 \over r^2+l^2} .\eqno(23)$$
 Comparing this with (15) we will see that
 $$8\pi \eta^2=2k \eqno(24)$$
 which gives the constant $k$  in terms of the vacuum value of the 
scalar field.\\
It may also be noted that neither dual-Schwarzschild (13b) nor dual-NUT
(13c) are asymptotically flat.

\section{Geodesics}
 The metric of a  NUT space with a global monopole has almost the same
 form as
 the NUT metric without a global monopole. Indeed one can write (13b) in the
  form
 $$A = B^{-1} =(1-2k)\left[1-{2(Mr+l^2)\over (r^2+l^2)}\right]$$
 where now $M={m\over (1-2k)}$. All the
 geodesics of NUT space, including the null ones, lie on spatial cones
 (Lynden-Bell and Nouri-Zonoz 1998).
 Following the same approach one can show that all the
 geodesics of NUT space with global monopole charge  also lie on
 spatial cones with the semi-angle of the cone given by
 $${\rm tan}\chi={2l \varepsilon \over L}$$ where
 $$\varepsilon = A(\dot t -2l{\rm cos}\theta \dot\phi)$$
 and ``${\;}^{.}$'' represents the differentiation with respect to an
 affine
 parameter.
 Thus the global monopole only contributes through $A$ in the
semi-angle of the cone. Its effect can be seen more explicitly
 by neglecting the mass and NUT parameters in the metric. We  can then write
 $$ds^2=(1-2k)dt^2 - {dr^2\over (1-2k)}-r^2(d\theta^2+{\rm sin}^2
 d\phi^2)$$
 which upon rescaling of $t$ and $r$ reads
 $$ds^2=dT^2-dR^2 - \left[R^2(1-2k)\right](d\theta^2+{\rm sin}^2 d\phi^2)
.$$
 This metric is dual to flat spacetime which can also be looked upon
as a spacetime of uniform gravitational potential , 
$g_{00}=-g^{-1}_{11}=1+2\phi$ (Dadhich 1997,1997b). It describes
a space with a
deficit solid angle in which
 the area of a sphere of radius $R$ is not $4\pi R^2$, but
$4\pi(1-2k)R^2$.
 The surface $\theta = \pi/2$ has the geometry of a cone with the deficit
 angle
 $\Delta \phi = 2\pi k$ $(k\ll 1)$. When a global monopole is added to a
Schwarzschild black hole both the particle orbits and the Hawking
radiation are merely rescaled (Dadhich et al 1998d).
 \section {Discussion}
 Using the decomposition of Riemann curvature with respect to a unit time
 like
 vector into electric and magnetic components we have defined a duality
 transformation between passive and active {\it electric} parts of the
 field. Though the vacuum equation was in general
invariant under the duality transformation, yet it is possible to construct
solutions dual to the well-known stationary solutions. This is
because there was a free equation in the vacuum set which did not
participate in determining the solution and hence could be
tampered with suitably to break the duality symmetry (11) of the vacuum
equation. The tampering would not however affect the vacuum solutions and would
lead to distinct daul solutions which remarkably imbibe a global monopole.
This is an interesting property of the field which is uncovered by the
appropriate modification of the vacuum equation (Dadhich 1997). \\
Like the massless global monopole with $m = l = 0$ in the solution (13),
non-static global texture spacetime with the equation of state $\rho + 3p
= 0$ can be shown to be dual-flat (Dadhich 1997). Global monopoles and
global textures which result from global symmetry beraking are stable
topological defects. The most intrguing feature of the duality
transformation is that it generates these topological defects in the
original Einstein solution. In spherical symmetry it is possible to give a
general prescription (Dadhich \& Patel 1998b) for writing solution dual
to any solution. Applications of the topological defects in cosmology have
been considered extensively (Vilenkin and Shellard, 1994). Does the
association of topological defects
with the duality transformation indicate a manifestation of something
deeper?
\section*{Acknowledgements}
We thank one of the referees for introducing Letelier's interpretation of
metric (13b) to us.  D.L-B and M.N-Z thank the members of IUCAA, Pune,
for their warm hospitality during their stay there. 
 M.N-Z acknowledges the support of the Ministry of Culture and 
Higher Education of Iran. D.L-B is a PPARC Senior Fellow.


\begin{thebibliography}{}
\bibitem {} Barriola, M. and Vilenkin, A. (1989), Phys.Rev.Lett., 63, 341.
\bibitem {} Dadhich, N. (1997), On Electrogravity duality, gr-qc/9712021
submitted to GRG.
\bibitem {} Dadhich, N. (1997b), On the Schwarzschild field, gr-qc/9704068.

\bibitem {} Dadhich, N. \& Patel, L. K. (1998a), Dual to Kerr solution, to be
submitted.
\bibitem {} Dadhich, N. \& Patel, L. K. (1998b), On spacetimes 
dual to spherically symmetric spacetimes, submitted.
\bibitem {} Dadhich, N., Patel, L. K. \&  Tikekar, R.  (1998c)
 Class. Quantum Grav. 15, L27.
\bibitem {} Dadhich, N., Narayan, K. \&  Yajnik, U. (1998d),
Pramana, 50, 307, gr-qc/9703034.
\bibitem {} Demiansky, M., Newman E. T., Bulletin De l'academie Polo
 naise des Sciences, Serie des Sciences math., astr. et phys.-Vol. XIV,
  No.11, 1966.
\bibitem {} Ehlers, J., Colloques internationaux C.N.R.S. No. 91
( Les theories relativistes de la gravitation), 275, 1962.
\bibitem {} Geroch, R., J.Math.Phys., 12, 918, 1971.

\bibitem {} Letelier, P. S., Phys.Rev.D, 6, 1294 (1979). 

\bibitem {} Lynden-Bell, D., Nouri-Zonoz, M., Rev. Mod. Phys., Vol. 70,
No. 2, April 1998.
\bibitem {} Misner, C.W., Thorne, K. S. and Wheeler, J. A., Gravitation,
Freeman, 1973.
\bibitem {} Vilenkin, A.  and  Shellard, E. P. S., Cosmic strings and other
topological defects, CUP, 1994.
\end{thebibliography}
\end{document}